\documentclass[10pt,letterpaper]{article}
\usepackage{opex3}

\usepackage[dvips]{epsfig}
\usepackage{subfigure}
\usepackage{float}
\usepackage{amsmath}
\usepackage{amssymb}
\usepackage{amsfonts}
\usepackage{rotating}
\usepackage{euscript}
\usepackage{subfigure}
\usepackage{enumerate}
\usepackage{hhline}
\usepackage{threeparttable}
\usepackage{supertabular}
\usepackage{pslatex}
\usepackage{tabularx}
\usepackage{color}
\usepackage{multirow}

\newcommand{\vecb}[1]{\mathbf{#1}}

\newcolumntype{Y}{>{\centering\arraybackslash}X}

\begin{document}

\title{Optical and mechanical design of a ``zipper'' photonic crystal optomechanical cavity}

\author{Jasper Chan, Matt Eichenfield, Ryan Camacho, and Oskar Painter}

\address{Thomas J. Watson, Sr., Laboratory of Applied Physics, California Institute of Technology, \\ Pasadena,
CA 91125}

\email{opainter@caltech.edu} 

\homepage{http://copilot.caltech.edu} 


\begin{abstract}
Design of a doubly-clamped beam structure capable of localizing mechanical and optical energy at the nanoscale is presented.  The optical design is based upon photonic crystal concepts in which patterning of a nanoscale-cross-section beam can result in strong optical localization to an effective optical mode volume of $0.2$ cubic wavelengths $\left((\lambda_{c})^3\right)$.  By placing two identical nanobeams within the near field of each other, strong optomechanical coupling can be realized for differential motion between the beams.  Current designs for thin film silicon nitride beams at a wavelength of $\lambda=1.5$ $\mu$m indicate that such structures can simultaneously realize an optical $Q$-factor of $7 \times 10^6$, motional mass $m_{u} \sim 40$ picograms, mechanical mode frequency $\Omega_{M}/2\pi \sim 170$ MHz, and an optomechanical coupling factor ($g_{\text{OM}} \equiv \text{d}\omega_{c}/\text{d}x = \omega_{c}/L_{\text{OM}}$) with effective length $L_{\text{OM}} \sim \lambda = 1.5$ $\mu$m.
\end{abstract}




\section{Introduction}

At a macroscopic level, the interaction of light with the mechanical degrees of freedom of a dielectric object can be calculated by considering the flux of momentum into or out of an object using the Maxwell stress-energy tensor.  At a microscopic level, as in the case of atomic physics, one can define an interaction Hamiltonian between an atom and the light field in order to derive the various mechanical forces on the atom's center of mass, which in general depends upon both the external and internal degrees of freedom of the atom\cite{ref:Metcalf_van_der_Straten}.  In the case of a dielectric mechanical resonator, a direct relationship between the macroscopic dielectric and microscopic atomic theories can be made, and useful analogies may be forged\cite{KippenbergOE,ref:Kippenberg_Sc_review}.  The interactions of light with mechanically resonant objects is currently being actively explored in the field of cavity optomechanics as a means to obtain ground-state cooling of a macroscopic mechanical resonator\cite{ref:Kippenberg_Sc_review,ref:Arcizet,ref:Gigan1,ref:Schliesser2,ref:Corbitt1,ref:ThompsonJD1,ref:Kippenberg4,ref:Regal1}.  The strength of optomechanical interactions in these system can be quantified on a per-photon basis by the rate of change of the cavity resonance frequency ($\omega_{c}$) with mechanical displacement amplitude ($u$), $g_{\text{OM}} \equiv \text{d}\omega_{c}/\text{d}u = \omega_{c}/L_{\text{OM}}$.  $L_{\text{OM}}$ is an effective length over which a cavity photon's momentum can be exchanged with the mechanical system. In this work we describe a simple doubly clamped nanobeam system (a so-called \emph{zipper} cavity) which allows for the combined localization of optical and mechanical energy in a nanoscale structure so as to provide extremely large optomechanical coupling due to the gradient optical force.  Optical energy is localized within the center of the cantilever using a one-dimensional photonic crystal in combination with total internal reflection.  Beyond the analysis provided here, in the future, optimization of both the optical and mechanical properties of these chip-based structures should allow for a variety of new applications from precision metrology\cite{ref:Stowe1} to tunable photonics\cite{ref:Povinelli1,ref:Notomi4}.

The outline for the paper is as follows.  We begin with the optical design of a one-dimensional photonic crystal in a siliocn nitride nanobeam.  Finite-element-method electromagnetic simulations are used to deduce the level of optical localization and the relevant optical losses within the struture.   The mechanical properties of the zipper cavity are studied next, with numerical simulations used to determine the lower-lying mechanical eigenmodes.  The tuning properties of a double nanobeam photonic crystal are then computed to estimate the strength of the optomechanical coupling for the differential in-plane motion of the beams.  We conclude with a comparison of the zipper cavity properties with other more macroscopic optomechanical systems, and a discussion of the future prospects for these sorts of chip-based gradient optical force devices.

\section{Optical design and simulation}

\begin{figure*}[ht]
\begin{center}
\includegraphics[width=1\columnwidth]{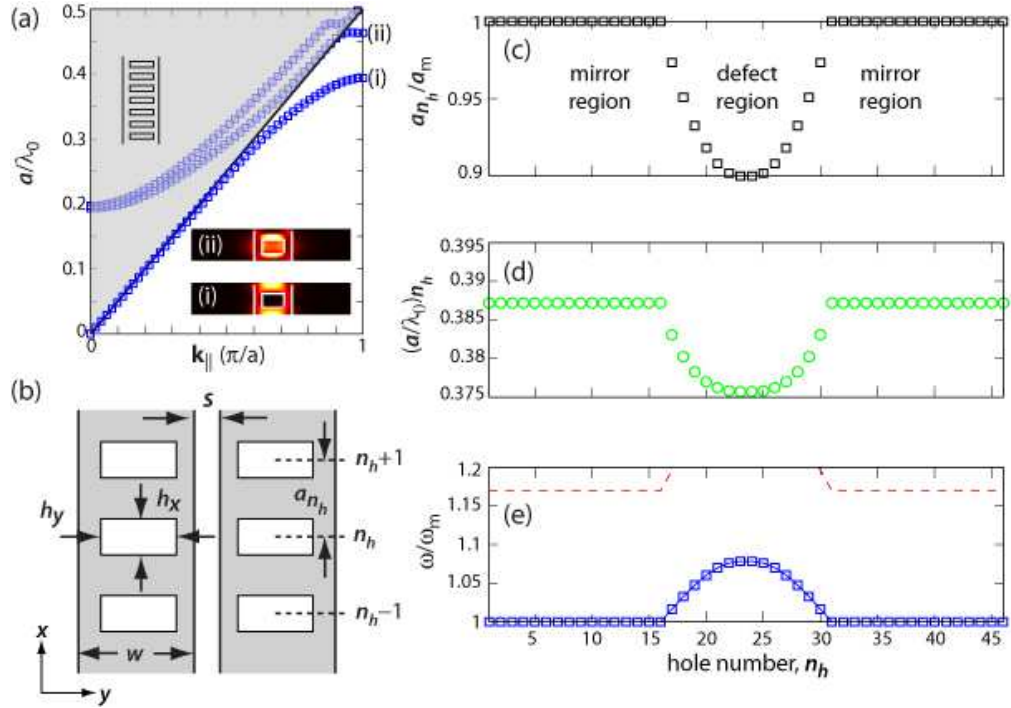}
\caption{Bandstructure properties of the photonic crystal nanowire. (a) Axial bandstructure of the single beam photonic crystal structure, with nominal width and thickness.  $\vecb{k}_{||}$ is the wavevector of light in the direction of the 1D photonic lattice, $a$ is the lattice period, and $a/\lambda_{0}$ is the ``normalized'' optical frequency.  The light cone, denoted by the grey area and deliniated by the black light line, represents regions of frequency-wave-vector space in the bandstructure diagram in which light can radiate into the two transverse directions orthogonal to the axis of the photonic lattice.  The two inset images show the electric field energy density of the valence (i) and conduction (ii) band-edge modes (the white outline is a contour plot of the refractive index of the nanowire).  (b) Schematic of the double beam zipper cavity indicating the slot gap ($s$), the lattice constant ($a$), the beam width ($w$), and the axial and transverse hole lengths, $h_{x}$ and $h_{y}$, respectively.  (c) Lattice constant (normalized to the lattice period in the mirror section of the cavity, $a_{m}$) versus hole number ($n_{h}$) within the photonic lattice of the cavity.  (c,d) Resulting frequency of the valence band-edge mode versus hole number.  In (d) the normalized frequency in terms of the local lattice constant, $(a/\lambda_{0})_{n_{h}}$, is displayed.  In (e) the local band-edge frequency is referenced to the valence band-edge in the mirror section of the cavity.  The solid blue (dashed red) curve is the valence (conduction) band-edge.}
\label{fig:single_beam}
\end{center}
\end{figure*}

The optical design of the zipper cavity utilizes a quasi-1D photonic crystal structure to localize optical modes to the center of a nanoscale cross-section beam.  There are a number of different, but related, design methodologies used to form low-loss optical resonances in such photonic crystal ``nanowires''\cite{ref:Foresi,ref:Sauvan1,ref:Zain1,ref:Velha2,ref:Notomi7,ref:McCutcheonM1}.  In this paper we utilize concepts based upon an envelope picture of the guided photonic crystal modes\cite{ref:Painter14} in the patterned nanobeam.  We choose to work with ``acceptor''-type modes formed from the lower-lying band of modes (the so-called ``valence'' band in reference to electronic semiconductor crystals) at the Brillouin zone boundary.  In this way, the mode frequencies are as far away from the light line as possible, reducing leakage into the surrounding air-cladding of the nanobeam.  Owing to the optically thin photonic wires considered in this work, the modes of predominantly in-plane polarization (TE-like modes) are most strongly guided and are the modes of primary interest here.  In order to form acceptor modes, the bandedge of the valence band must be increased near the center of the cavity, and decreased in the mirror sections defining the end of the cavity.  This is a result of the negative group velocity dispersion of the valence band modes\cite{ref:Painter14}.  

We have chosen to perform designs based upon thin films of silicon nitride, as opposed to higher refractive index materials such as silicon-on-insulator (SOI), for several reasons.  One reason is that silicon nitride can be grown on silicon wafers with very high optical quality across a wide range of wavelengths covering the visible to the mid-infrared. We have measured\cite{ref:Barclay8} optical $Q$-factors in excess of $3 \times 10^{6}$ for whispering-gallery modes of microdisks formed from stoichiometric silicon nitride deposited by low-pressure-chemical-vapor-deposition (LPCVD).  A second reason is that LPCVD-deposited stoichiometric silicon nitride films on silicon have a large internal tensile stress, which has been shown to be critical in producing high-$Q$ mechanical resonances in doubly-clamped nanobeams\cite{ref:Verbridge1,ref:Verbridge3}.  An additional concern is the two-photon absorption present in smaller bandgap semiconductors such as silicon and gallium arsenide, which results in additional free-carrier absorption (FCA), and which can result in significant parasitic effects such as thermo-optic and free-carrier dispersion in small-volume photonic crystal nanocavities\cite{ref:Barclay7}.  Silicon nitride, with its large bangap ($\sim 3$ eV), requires three (as opposed to two) $1$ eV photons to be absorbed simultaneously, greatly reducing nonlinear absorption in the near-IR. An obvious drawback of using silicon nitride thin films, is the lower refractive index ($n \sim 2$) of these films in comparison to semiconductor films ($n \sim 3.4$).  As is shown below (and in Ref. \cite{ref:McCutcheonM1}), with carefully chosen designs, high-$Q$ photonic crystal optical cavities of sub-cubic-wavelength mode volume can still be formed in silicon nitride thin films.   

\begin{figure*}[ht]
\begin{center}
\includegraphics[width=1\columnwidth]{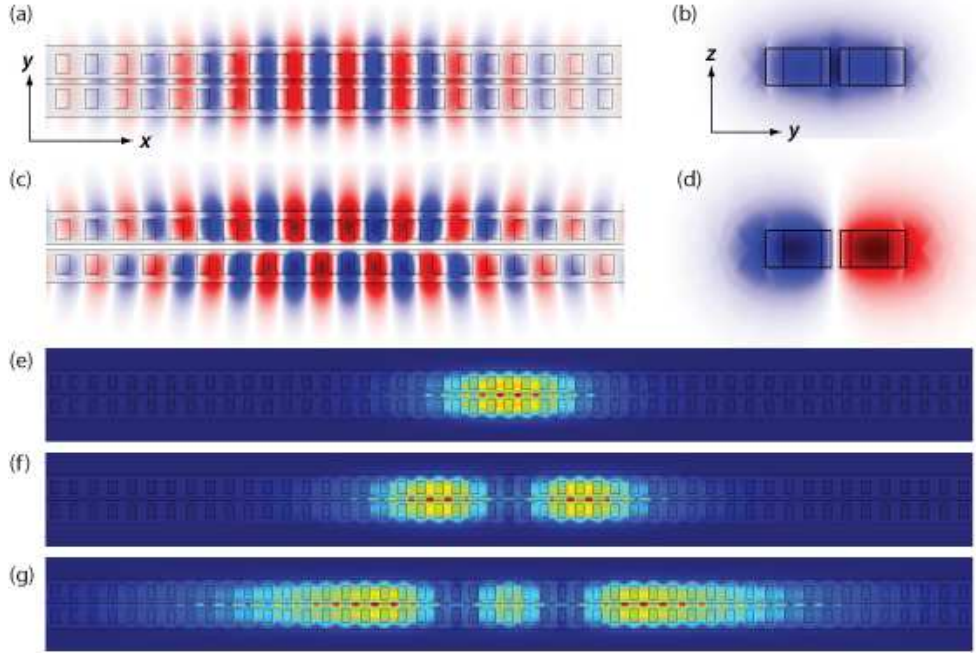}
\caption{Transverse electric field ($\vecb{E}_{y}$) mode profile of the fundamental bonded mode (TE$_{+,0}$):  (a) top-view, (b) cross-section.  Transverse electric field ($\vecb{E}_{y}$) mode profile of the fundamental anti-bonded mode (TE$_{-,0}$):  (c) top-view, (d) cross-section.  The field colormap corresponds to +1 (red), 0 (white), and -1 (blue).  Electric field energy density of the (e) fundamental (TE$_{+,0}$), (f) second-order (TE$_{+,1}$), and (g) third-order (TE$_{+,2}$) bonded optical modes.  The intensity colormap ranges from  +1 (red) to 0 (blue).}
\label{fig:optical_mode_profiles}
\end{center}
\end{figure*}

The bandstructure of a single beam silicon nitride nanowire, calculated using the MIT photonic bands software package\cite{ref:MPB_note}, is displayed in Fig. \ref{fig:single_beam}.  As described in more detail below, the simulation is performed for a ``nominal'' structure defined by a lattice normalized beam thickness ($\bar{t}=t/a=2/3$) and beam width ($\bar{w}=w/a=7/6$).  The refractive index of the silicon nitride beam is taken as $n=2$, and a resolution of $64$ points per axial lattice period is used to ensure accurate band frequencies.  Our coordinate convention is (see Fig. \ref{fig:optical_mode_profiles}): (i) $x$ the in-plane coordinate along the long axis of the cavity, (ii) $y$ the in-plane transverse coordinate, and (iii) $z$ the out-of-plane transverse coordinate.  Only the lower-lying bands with modes of even parity in the $\hat{z}$-direction and odd parity in the $\hat{y}$-direction are shown, corresponding to the fundamental TE-like modes of the beam waveguide.  As indicated by the electric field energy density plots of the two lowest lying band-edge modes (inset to Fig. \ref{fig:single_beam}(a)), the valence band lies predominantly in the region of the high refractive index silicon nitride beam, whereas the upper ``conduction'' band mode lies predominantly in the region of the air hole patterning.  

The formation of localized optical cavity resonances is accomplished by introducing a ``defect'' into the photonic lattice.  The defect region in the structures studied here consists of a quadratic grade in the lattice constant of the linear array of air holes near the cavity center.  In order to reduce transverse radiation loss, we choose to use a defect which supports an odd symmetry fundamental mode along the axial direction.  Since the valence band modes tend to have electric field intensity predominantly inside the high-dielectric region (and nodes of the electric field in the low-dielectric constant air holes), a defect in which an air hole is at the center of the cavity yields an odd parity fundamental mode along the axial direction.  Here, and in what follows, we use a cavity defect region consisting of the central $15$ holes, with the lattice period varied from a nominal value in the outer \emph{mirror} section ($a_{m}$) to $90 \%$ of the nominal value at the center of the defect region.  This was found to provide a good balance between axial-localization of the cavity modes and radiation loss into the $y$-$z$ transverse directions of the nanobeams.  In Fig. \ref{fig:single_beam}(c) we plot the \emph{local} lattice period, defined as $a_{n_{h}}=x(n_{h}+1) - x(n_{h})$, versus air hole number $n_{h}$ along the length of the cavity.  In Fig. \ref{fig:single_beam}(d) we plot the corresponding lattice-normalized frequency of the local TE-like valence band-edge modes of the single beam cavity.  The small variation in lattice-normalized frequency is a result of the distortion in the aspect ratio of the (perfectly periodic) structure as the lattice period is changed.  More useful is the plot in Fig. \ref{fig:single_beam}(e) which shows the frequency of the local valence band-edge (blue solid curve) and conduction band-edge (red dashed curve) modes normalized to the valence band-edge mode frequency in the mirror section of the cavity.  The quadratic grade in lattice constant results in a nearly-harmonic shift in the valence band-edge frequency versus position in the center of the cavity, with the band-edge frequency at the cavity center lying approximately mid-gap between the valence and conduction band-edges in the mirror section of the cavity.  

\begin{figure}[ht]
\begin{center}
\includegraphics[width=0.7\columnwidth]{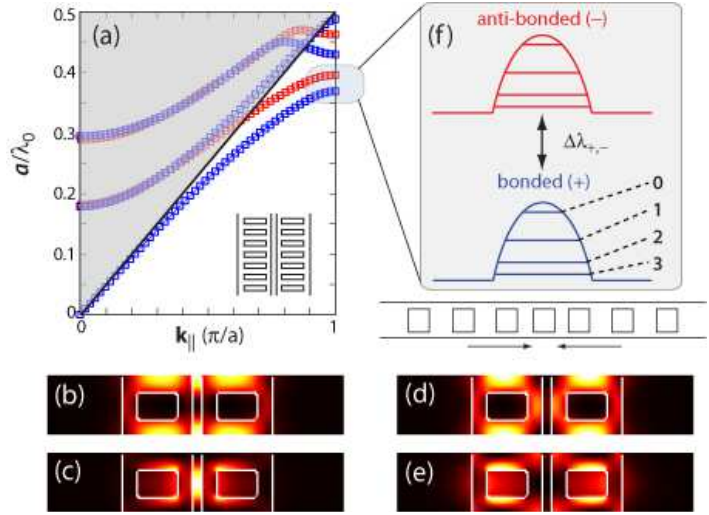}
\caption{Optical design principle of the zipper cavity. (a) Axial bandstructure of the double beam quasi-1D photonic crystal structure, with nominal width, thickness, and slot gap.  The blue curves are the bonded bands and the red curves are the anti-bonded bands.  Valence and conduction band-edge modes of the bonded (b,c) and anti-bonded (d,e) bands, respectively.  (f) Illustration of the defect cavity formation at the Brillouin-zone boundary.  The splitting between the two manifolds is indicated by $\Delta\lambda_{+,-}$.}
\label{fig:optical_design_schem}
\end{center}
\end{figure}

For the double-beam design of the zipper cavity, two photonic crystal nanowires are placed in the near-field of each other as illustrated in Fig. \ref{fig:single_beam}(b).  The strong coupling between two nearly-identical nanobeams results in a bandstructure consisting of even and odd parity superpositions of the TE-polarized single beam photonic bands (Fig. \ref{fig:optical_design_schem}(a)).  We term the even parity supermodes, \emph{bonded} modes, and the odd parity supermodes, \emph{anti-bonded} modes\cite{ref:Povinelli1}.  The anti-bonded manifold of resonant modes are shifted to higher frequency than the bonded manifold of modes, with splitting ($\Delta\lambda_{+,-}$) being dependent upon the slot gap ($s$) between the nanobeams.  The electric field energy density plots of the bonded valence and conduction band-edge modes are shown in Figs. \ref{fig:optical_design_schem}(b) and \ref{fig:optical_design_schem}(c), respectively.  The corresponding anti-bonded modes, with noticeably reduced energy density in the slot gap, are displayed in Fig. \ref{fig:optical_design_schem}(d,e).   Owing to the negative curvature of the valence band forming the localized cavity modes, the fundamental mode of the cavity for each manifold has the highest frequency, with the higher-order modes having reduced frequencies (schematically illustrated in Fig. \ref{fig:optical_design_schem}(f)).  In what follows, we will primarily be interested in the bonded mode manifold due to the larger electric field intensity of these modes in the slot gap.

With the zipper cavity now defined, we perform a series of finite-element-method (FEM), fully vectorial, 3D simulations of the localized cavity modes\cite{ref:COMSOL_note}.  In these simulations we use the graded cavity design described above.  The numerical mesh density is adjusted to obtain convergent values for the frequency and optical $Q$-factor of the cavity modes, and the structure is simulated with all available symmetries taken into account allowing for 1/8th the simulation volume.  Scattering boundary conditions are used in the outer axial and transverse boundaries to provide a nearly-reflectionless boundary for out-going radiation.  In the transverse direction ($y$-$z$) a cylindrical outer boundary is also used to reduce the total simulation volume.  Finally, a check of the accuracy of our FEM simulations was also performed through a series of equivalent simulations using finite-difference time-domain code (Lumerical\cite{ref:Lumerical_note}).  Figures \ref{fig:optical_mode_profiles}(a) and \ref{fig:optical_mode_profiles}(c) display the FEM-calculated electric field mode profiles of the fundamental bonded (TE$_{+,0}$) and anti-bonded (TE$_{-,0}$) modes, respectively, of the double beam zipper cavity.  Cross-sectional electric field profiles, displayed in Figs. \ref{fig:optical_mode_profiles}(b,d), clearly show the even and odd parity of the modes.  Also shown in Figs. \ref{fig:optical_mode_profiles}(e-g) are the electric field intensity of the lowest three bonded cavity mode orders, TE$_{+,0}$, TE$_{+,1}$, and TE$_{+,2}$.   

Design variations of the zipper cavity are performed around a ``nominal'' structure with the following (normalized) dimenensions: (i) beam width, $\bar{w} \equiv w/a_{m} = 700/600$, (ii) beam thickness, $\bar{t} \equiv t/a_{m} = 400/600$, (iii) axial hole length, $\bar{h}_{x} \equiv h_{x}/a_{m} = 267/600$, (iv) axial hole width, $\bar{h}_{y} \equiv h_{y}/a_{m} = 400/600$, and (v) slot gap, $\bar{s} \equiv s/a_{m} = 100/600$, where $a_{m}$ is the lattice periodicity in the cavity mirror section.  The length of the beam is set by the number of air hole periods in the cavity, which for the nominal structure is $N_{h}=47$ (23 holes to the left and right of the central hole, with the central 15 holes defining the defect region).  The filling fraction of the air holes in the nominal structure is $f = 25.4 \%$.  For a fundmamental bonded mode wavelength of $\lambda \sim 1500$ nm, the nominal lattice constant is $a_{m}=600$ nm, hence the normalization by $600$ in the above expressions for normalized dimensions. 

\begin{figure}[ht]
\begin{center}
\includegraphics[width=0.75\columnwidth]{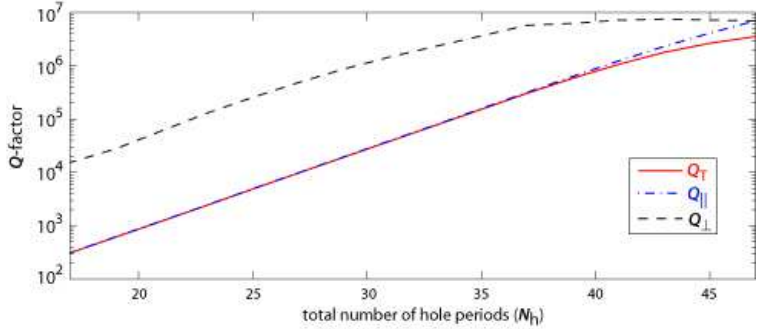}
\caption{Optical $Q$-factor (axial, transverse, and total) versus number of hole periods in the cavity, $N_{h}$.  The nominal structure corresponds to the maximum hole number in this plot, $N_{h}=47$.}
\label{fig:Q_Nh}
\end{center}
\end{figure}

Figure \ref{fig:Q_Nh} shows the simulated $Q$-factor of the fundamental bonded cavity mode (TE$_{+,0}$) versus the total number of periods $N_{h}$ of the cavity for the nominal structure.  The normalized cavity resonance frequency is calculated to be $a_{m}/\lambda_{c} = 0.3927$.  The radiation loss from the cavity is broken into two parts, yielding two effective $Q$-factors: (i) the axial radiation loss out the ends of the nanobeams (yielding $Q_{||}$), and (ii) the radiation loss transverse to the long axis of the zipper cavity and intercepted by the transverse boundary (yielding $Q_{\perp}$).  In Figure \ref{fig:Q_Nh}, $Q_{||}$ is seen to rise exponentially as a function of $N_{h}$, with an order of magnitude increase in $Q$-factor for every 6 additional periods of air holes.  The transverse $Q$ is seen to rise initially with hole periods, but then levels off and saturates at a value of $Q_{\perp} = 7 \times 10^6$.  The variation in $Q_{\perp}$ for structures with small $N_{h}$, and low $Q_{||}$, is a result of weak coupling between radiation loss into these (arbitrarily chosen) directions.  Small reflections at the end of the nanobeams results in a small amount of axial radiation making its way to the transverse boundary.  Nevertheless, a structure with $N_{h} > 47$ results in a total radiation $Q$-factor limited by the transverse $Q$-factor of $7 \times 10^6$.  This value is very large given the modest refractive index of the silicon nitride film and small cross-section of the nanobeams (large air filling fraction).

\begin{figure}[ht]
\begin{center}
\includegraphics[width=0.75\columnwidth]{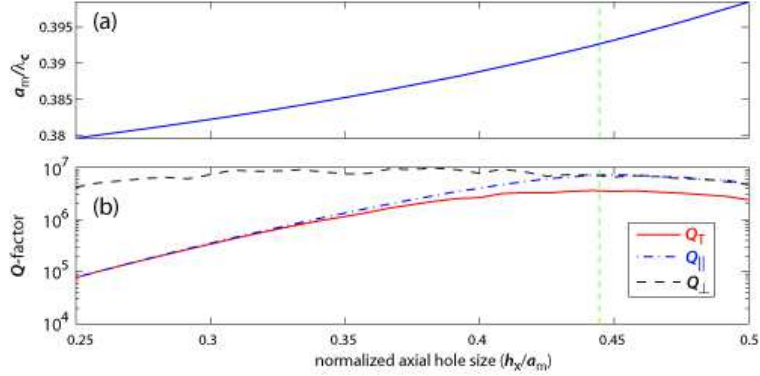}
\caption{(a) Normalized frequency and (b) Optical $Q$-factor (axial, transverse, and total) versus normalized axial hole length, $\bar{h}_{x}$.  The nominal structure is indicated by the dashed green line.}
\label{fig:Q_hx}
\end{center}
\end{figure}

\begin{figure}[ht]
\begin{center}
\includegraphics[width=0.75\columnwidth]{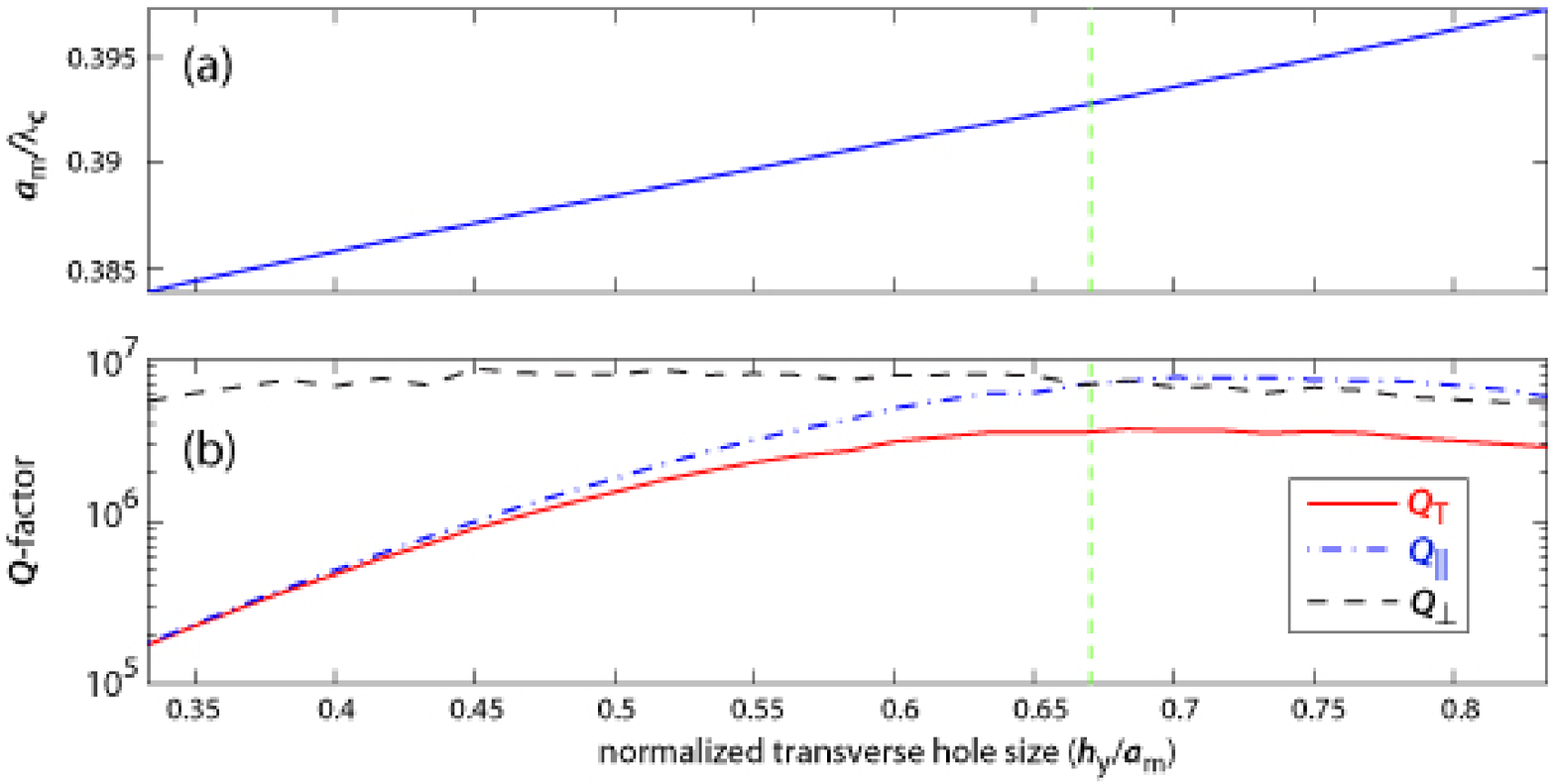}
\caption{(a) Normalized frequency and (b) Optical $Q$-factor (axial, transverse, and total) versus normalized transverse hole length, $\bar{h}_{y}$.  The nominal structure is indicated by the dashed green line.}
\label{fig:Q_hy}
\end{center}
\end{figure}

In order to study the dependence of $Q$-factor on the hole size, we have also simulated the nominal structure with varying axial and transverse hole size, as shown in Figs. \ref{fig:Q_hx} and \ref{fig:Q_hy}.  In each of the plots the axial $Q$-factor increases with increasing hole size, but then saturates, as the hole size approaches that of the nominal structure.  For hole sizes larger than the nominal structure, the $Q_{||}$ slightly drops, as does the transverse $Q$-factor.  This drop in transverse $Q$-factor is a result of the increased normalized frequency of the mode (higher air filling fraction), which pushes the mode closer to the air-cladding light line, increasing radiation into the cladding (the drop in $Q_{||}$ is a result of low-angle transverse radiation making its way to the boundary at the end of the nanobeams).  Therefore, the nominal structure represents an optimal structure in this regard, allowing for tight axial localization of the mode, without decreasing the transverse $Q$-factor.   

\begin{figure}[ht]
\begin{center}
\includegraphics[width=0.75\columnwidth]{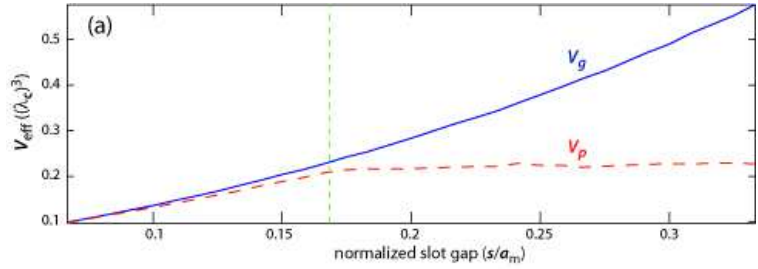}
\caption{Effective mode volume versus normalized slot gap, $\bar{s}$.  The nominal structure is indicated by the dashed green line.}
\label{fig:V_eff_s}
\end{center}
\end{figure}

The strength of light-matter interactions depend upon a position-dependent effective optical mode volume of a resonant cavity,

\begin{equation}
\label{eq:mode_v}
V_{\text{eff}}(\vecb{r}_{0}) \equiv \frac{\int \varepsilon(\vecb{r})|\vecb{E}(\vecb{r})|^2 \text{d}^3 \vecb{r}}{\varepsilon(\vecb{r}_{0})|\vecb{E}(\vecb{r}_{0})|^2},
\end{equation}  

\noindent where $\varepsilon(\vecb{r})$ is the position dependent dielectric constant and $\vecb{r}_{0}$ is the position of interest.  In the case of the zipper cavity, the strength of the optomechanical coupling depends upon $V_{\text{eff}}(\vecb{r}_{0})$ evaluated in the slot gap of the nanobeams.  Similarly, in the field of cavity-QED, the effective mode volume can be used to estimate the coherent coupling rate between an ``atom'' and the cavity field.  In Fig. \ref{fig:V_eff_s} we plot the effective mode volume versus slot gap size, $s$, between the nanobeams for the TE$_{0,+}$ mode of the nominal structure.  We plot two different effective mode volumes: (i) $V_{g}$, the effective mode volume evaluated at the center of the nanobeam gap near the center of the cavity where the field is most intense, and (ii) $V_{p}$, the minimum effective mode volume evaluated at the position of peak electric field energy density in the cavity.  For normalized gap widths, $\bar{s} < 1/6$ the modes of the nanobeam are strongly coupled resulting in a peak electric field intensity in between the slot gap (hence the two effective mode volumes track each other).  The effective mode volumes approach a value of $V_{\text{eff}} = 0.1 (\lambda_{c})^3$ as the slot gap approaches zero.  This small value is a result of the discontinuity in the dominant polarization of the TE mode ($E_{y}$) as it crosses into the slot gap\cite{ref:Robinson1}.  For slot gaps $\bar{s} > 1/6$, the minimum effective mode volume saturates at a value of $V_{p} = 0.225 (\lambda_{c})^3$ corresponding to that of a single nanobeam (i.e., no enhancement from energy being pushed into the slot gap).  The effective mode volume evaluated at the center of the slot gap, on the otherhand, continues to rise with slot gap due to the exponential decay of the field in the gap.  

\begin{figure}[ht]
\begin{center}
\includegraphics[width=0.75\columnwidth]{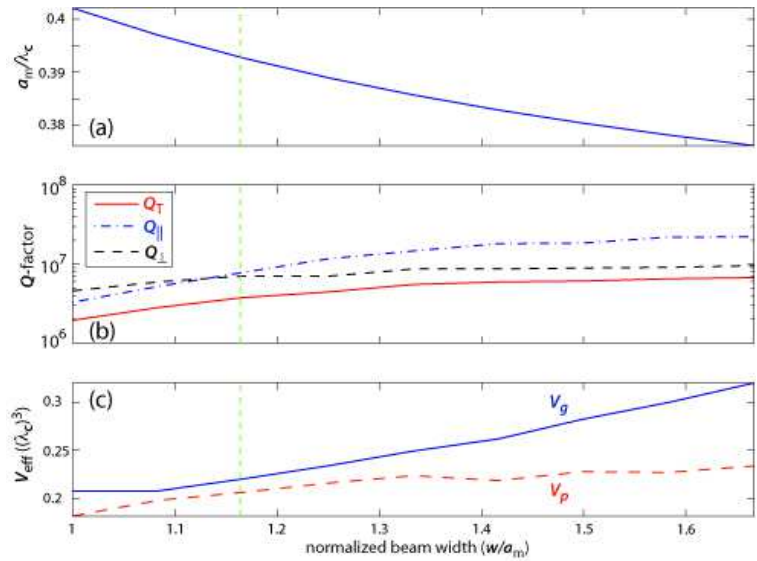}
\caption{(a) Normalized frequency, (b) $Q$-factor and (c) Effective mode volume versus normalized beam width.  The nominal structure is indicated by the dashed green line.}
\label{fig:beam_width}
\end{center}
\end{figure}

A final variation considered is the nanobeam width.  In Fig. \ref{fig:beam_width} the nominal structure is varied by adjusting the normalized beam width while holding the filling fraction of the air holes fixed.  This is done by scaling the the transverse hole length with the beam width.  The axial $Q$-factor increases with the beam width due to the increased effective index of the guided mode, and thus increased effecive contrast of the quasi-1D photonic crystal.  Due to the reduced lateral localization of the cavity mode, the effective mode volumes also increase with beam width. In contrast, the transverse $Q$-factor remains approximately constant.  

\section{Mechanical mode analysis}

\begin{figure*}[ht]
\begin{center}
\includegraphics[width=1\columnwidth]{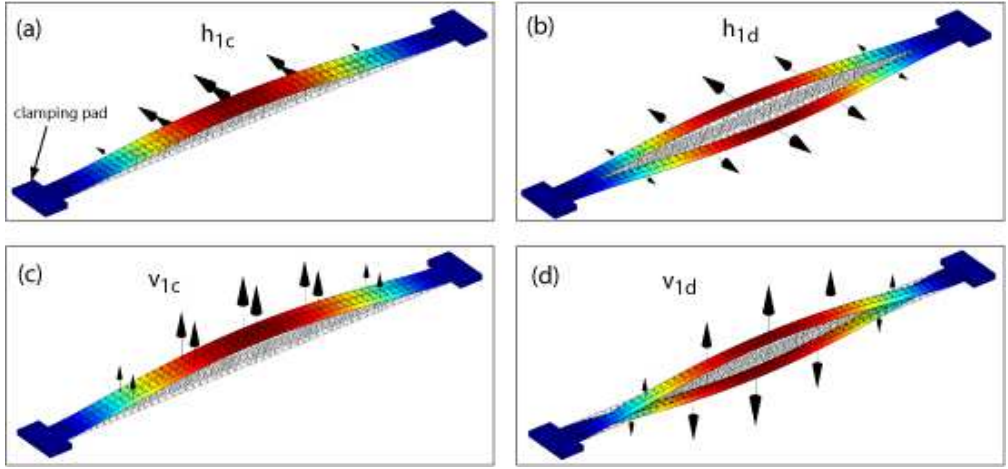}
\caption{Mechanical eigenmode displacement plots: (a) 1st-order in-plane common mode, (b) 1st-order in-plane differential mode, (c) 1st-order out-of-plane common mode, and (d) 1st-order out-of-plane differential mode.  The color represents total displacement amplitude, as does the deformation of the structure.  The arrows indicate the local direction of displacement.}
\label{fig:mech_modes}
\end{center}
\end{figure*}

The mechanical modes of the zipper cavity can be categorized into common and differential modes of in-plane (labeled $h$) and out-of-plane motion (labeled $v$) of the two nanobeams.  In addition there are compression (labeled $c$) and twisting (labeled $t$) modes of the beams.  In this work we focus on the in-plane differential modes, $h_{qd}$, as these modes are the most strongly opto-mechanically coupled due to the large change in the slot gap per unit (strain) energy.  We use FEM numerical simulations to calculate the mechanical mode patterns and mode frequencies, the first few orders of which are shown in Fig. \ref{fig:mech_modes}.  The material properties of silicon nitride for the FEM simulations were obtained from a number of references.  Where possible we have used parameters most closely associate with stoichiometric, low-pressure-chemical-vapor-deposition (LPCVD) silicon nitride deposited on $\langle 100 \rangle$ Si: mass density $\rho = 3100$ kg/m$^3$, Young's modulus $Y \sim 290$ GPa, internal tensile stress $\sigma \sim 1$ GPa\cite{ref:Verbridge1,ref:Verbridge3}, coefficient of thermal expansion $\eta_{TE} = 3.3 \times 10^{-6}$ K$^{-1}$, thermal conductivity $\kappa_{th} \sim 20$ W/m/K, and specific heat $c_{sh} = 0.7$ J/g/K.  

For the mechanical mode properties tabulated in Table \ref{tab:mech_mode_properties} we have analyzed the nominal zipper cavity structure at an operating wavelength of $\lambda \approx 1.5$ $\mu$m, corresponding to a geometry with $a_{m}=600$ nm, $t=400$ nm, $s=100$ nm, $w=700$ nm, $h_{x}=267$ nm, and $h_{y}=400$ nm.  The total number of air holes is set to $N_{h}=55$, ensuring a theoretical $Q$-factor dominated by transverse radiation ($Q_{\perp}$), yielding a cavity length of $l=36$ $\mu$m.  The zipper cavity is clamped at both ends using a fixed boundary conditions at the far ends of the ``clamping pads'' seen in Fig. \ref{fig:mech_modes}.  This clamping scheme is suitable for estimating the mechanical mode eigenfrequencies, although more complex clamping schemes envisioned for real devices will likely introduce modified splittings betweeen nearly-degenrate common and differential modes.  The effective spring constant listed in Table \ref{tab:mech_mode_properties} for each mode is based upon a motional mass equal to that of the true physical mass of the patterned nanobeams, $m_{u} \approx 43$ picograms (see below for self-consistent definition of motional mass).  

The resulting frequency for the fundamental $h_{1d}$ mode is $\Omega/2\pi \approx 8$ MHz.  The in-plane mode frequency of a doubly clamped beam, with $l \gg w,t$, is approximately given by\cite{ref:Verbridge1}:

\begin{equation}
\label{eq:mech_freq}
\Omega_{q}/2\pi = \frac{q^2\pi}{2l^2}\sqrt{\frac{YI_{y}}{\rho A}}\sqrt{1+\frac{\sigma A l^2}{q^2YI_{y}\pi^2}},
\end{equation}

\noindent where $q$ is the mode index (approximately an integer), $A=tw$ is the cross-sectional area of the beams, and $I_{y} = t w^3/12$ is the cross-sectional moment of inertial about the in-plane axis ($\hat{y}$) of the beam.  This fits the numerical data reasonably accurately assuming an effective beam width of $w^{\prime} = (1-f) w$.  From the scaling in eq. (\ref{eq:mech_freq}), one finds that for mode number $q \geq 3$ (where internal stress can be neglected and $\Omega_{q}/2\pi \approx \frac{q^2\pi}{2l^2}\sqrt{\frac{YI_{y}}{\rho A}}$) that the in-plane frequency scales inversely with the square of the beam length, linearly with the beam width, and independent of the beam thickness.  Therefore, a linear increase in the resonant frequency can be obtained by moving to shorter optical wavelengths and scaling the structure with wavelength.  For much larger increases in mechanical resonance frequency, one must resort to higher-order in-plane modes of motion.  Optomechanical coupling to these modes is discussed below.

\renewcommand{\arraystretch}{1.1}
\renewcommand{\extrarowheight}{0pt}
\begin{table}[ht]
\caption{Summary of mechanical mode properties.}
\label{tab:mech_mode_properties}
\begin{center}
\begin{tabularx}{\linewidth}{YYYY}
\hline
\hline
Mode label & $\Omega_{M}/2\pi$ (MHz) & $k_{u}$ (N/m) & $g_{\text{OM}}$ ($\omega_{c}/\lambda_{c}$) \\
\hline
$v_{1c}$ & 5.95 & 60 & \\
$v_{1d}$ & 6.15 & 64 & \\
$v_{2c}$ & 12.3 & 257 & \\
$v_{2d}$ & 12.7 & 274 & \\
$v_{3c}$ & 19.2 & 626 & \\
$v_{3d}$ & 20.0 & 679 & \\
\hline
$h_{1d}$ & 7.91 & 106 & 1.24 \\
$h_{1c}$ & 7.94 & 107 & $\sim 0$ \\
$h_{2d}$ & 18.2 & 562 & $\sim 0$ \\
$h_{2c}$ & 18.3 & 568 & $\sim 0$ \\
$h_{3d}$ & 31.8 & 1717 & 1.16\\
$h_{3c}$ & 32.0 & 1738 & $\sim 0$ \\
$h_{9d}$ & 167.7 & $4.77 \times 10^4$ & 0.63 \\
$h_{9c}$ & 168.0 & $4.79 \times 10^4$ & $\sim 0$ \\
\hline
$t_{1d}$ & 41.0 & 2854 & \\
$t_{1c}$ & 41.1 & 2868 & \\
\hline
$c_{1c}$ & 78.6 & $1.05 \times 10^4$ & \\
$c_{1d}$ & 79.4 & $1.07 \times 10^4$ & \\
\hline
\hline
\end{tabularx}
\end{center}
\end{table}
\renewcommand{\arraystretch}{1.0}
\renewcommand{\extrarowheight}{0pt}

We have also studied the expected thermal properties of the zipper cavity, again assuming a $1.5$ $\mu$m wavelength of operation.  Due to the air-filling-fraction of the etched holes in the zipper cavity nanobeams, the thermal conductivity of the patterned beams is approximately $\Gamma_{th}=75 \%$ of the bulk value.  A simple estimate for the thermal resistance of the zipper cavity is $R_{th} \sim l/(8 t w \Gamma_{th}\kappa_{th}) \approx 1.15 \times 10^{6}$ K/W, where the factor of $1/8$ comes from the ability for heat to escape out either end of the nanobeams and in either direction.  The physical mass of the zipper cavity, taking into account the etched holes, is approximately $m = 43$ picograms.  The heat capacity of the zipper cavity is then $c_{h} \approx 3 \times 10^{-11}$ J/K.  From the heat capacity and the thermal resistance, the thermal decay rate is estimated to be $\gamma_{th} = 1/R_{th} c_{h} \sim 2.9 \times 10^4$ s$^{-1}$.  Finite-element-method simulations of the thermal properties of the zipper cavity yield an effective thermal resistance of $R_{th} = 1.09 \times 10^{6}$ K/W and a thermal decay rate of $\gamma_{th} = 5.26 \times 10^4$ s$^{-1}$ for temperature at the center of the zipper cavity, in reasonable correspondence to the estimated values.  

\section{Optomechanical coupling}

With the optical and mechanical modes of the zipper cavity now characterized, we proceed to consider the optomechanical coupling of the optical and mechanical degress of freedom.  As described at the outset, the parameter describing the strength of optomechanical coupling is the frequency shift in the cavity mode frequency versus mechanical displacement, $g_{\text{OM}} \equiv \text{d}\omega_{c}/\text{d}u$, where $\omega_{c}$ is the cavity resonance frequency and $u$ represents an amplitude of the mechanical displacement.  In the case of the commonly studied Fabry-Perot cavity\cite{ref:ThompsonJD1}, $g_{\text{OM}} = \omega_{c}/L_{c}$, where $L_{c}$ is approximately the physical length of the cavity.  A similar relation holds for whispering gallery structures, such as the recently studied microtoroid\cite{KippenbergOE}, in which the optomechanical coupling is proportional to the inverse of the radius of cavity ($R$), $g_{\text{OM}} = \omega_{c}/R$.  Both these devices utilize the radiation pressurce, or scattering, force of light.  By comparison, the zipper cavity operates using the gradient force for which the optomechanical coupling length can be on the scale of the wavelength of light, $L_{\text{OM}} \sim \lambda_{c}$.  Similar to the scaling found in the previous section for the effective mode volume, the optomechanical coupling length scales exponentially with the slot gap, $L_{\text{OM}} \sim \lambda_{c} e^{\alpha s}$, where $\alpha$ is proportional to the refractive index contrast between the nanobeams forming the zipper cavity and the surrounding cladding.

\begin{figure}[ht]
\begin{center}
\includegraphics[width=\columnwidth]{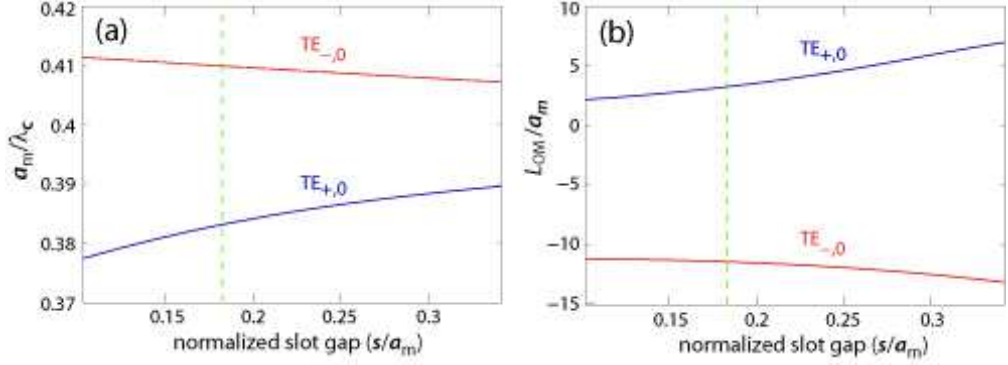}
\caption{(a) Bonded and anti-bonded mode tuning curves versus normalized nanobeam gap.  (b) Corresponding normalized effective optomehcanical coupling length, $\overline{L}_{\text{OM}} \equiv L_{\text{OM}}/a_{m}$.  The nominal structure is indicated by the dashed green line.}
\label{fig:optical_mode_tuning}
\end{center}
\end{figure}

In Fig. \ref{fig:optical_mode_tuning}(a) we plot the tuning curve for the nominal zipper cavity structure studied in the previous two sections versus the normalized slot gap width, $\bar{s} \equiv s/a_{m}$.  Due to the strong intensity of the bonded mode in the center of the slot gap, it tunes more quickly than the anti-bonded mode.  This tuning curve can be used to estimate the optomechanical coupling for the in-plane differential mode of motion of the zipper cavity nanobeams.  The in-plane common mechanical modes and both types of vertical mechanical modes are expected to provide a much smaller level of optomechanical coupling due to the reduced change in slot gap with these types of motion.  For complex geometries and motional patterns, one must use a consistent definition of displacement amplitude, $u$, in determing $g_{\text{OM}}$, $m_{\text{eff}}$ (motional mass), and $k_{\text{eff}}$ (effective spring constant).  In this work we use a convention in which $u(t)$ represents the amplitude of motion for a normalized mechanical eigenmode displacement field pattern:

\begin{equation}
\label{eq:motion_def}
\vecb{u}_{n}(\vecb{r},t) = u_{n}(t)\frac{\vecb{f}_{n}(x)}{\sqrt{\frac{1}{l}\int_0^{l}|\vecb{f}_{n}(x)|^2\text{d}x}},
\end{equation}

\noindent where $n$ is a mode label, $l$ is the length of nanobeam, and, for the simple beam geometry considered here, the displacement vector is only a function of position along the long axis of the nanobeams  ($x$).  With this definition of \emph{amplitude}, the effective motional mass is simply the total mass of the two nanobeams ($m_{u}=43$ picograms), and the effective spring constant is defined by the usual relation $k_{u} = m_{u}\Omega_{M}^2$, with $\Omega_{M}$ the mechanical eigenmode frequency.  The amplitude associated with zero-point motion and used in the equipartition theorem to determine the thermal excitation of the mechanical mode is then $u_{n}(t)$.  Note, an alternative, but equally effective method, defines the amplitude first, and then adjusts the effective motional mass based upon the strain energy of the mechanical motion.  

Our chosen normalization prescription yields (approximately) for the $q$th odd-order in-plane differential mechanical mode, $\vecb{u}^{\text{o}}_{h_{qd}}(z,t) \approx u_{h_{qd}}(t)\left(\hat{y}_{1}\cos(q\pi x/l) + \hat{y}_{2}\cos(q\pi x/l)\right)$, where $\hat{y}_{1}$ and $\hat{y}_{2}$ are (transverse) in-plane unit vectors associated with first and second nanobeams, respectively, and which point in opposite directions away from the center of the gap between the nanobeams.  The even-order modes are anti-symmetric about the long axis of the cavity and are given approximately by, $\vecb{u}^{\text{e}}_{h_{qd}}(z,t) \approx u_{h_{qd}}(t)\left(\hat{y}_{1}\sin(q\pi x/l) + \hat{y}_{2}\sin(q\pi x/l)\right)$.  To be consistent then, with this definition of mode displacement amplitude, $g_{\text{OM}}$ must be defined in terms of the rate of change of cavity frequency with respect to \emph{half} the change in slot gap ($g_{\text{OM}} \approx \text{d}\omega_{c}/\frac{1}{2}\text{d}\delta s$), as the amplitude $u_{h_{qd}}(t)$ corresponds to a (peak) change in slot gap of $2u_{h_{qd}}(t)$.  

Fig. \ref{fig:optical_mode_tuning}(b) plots the optomechanical coupling length for each of the bonded and anti-bonded fundamental modes from the derivative of their tuning curves in Fig. \ref{fig:optical_mode_tuning}(a).  This plot shows that for a normalized slot gap of $\bar{s} = 0.1$ (or roughly $s = 0.04\lambda_{c}$), the optomechanical coupling length to the fundamental bonded optical mode can be as small as $L_{\text{OM}}/a_{m} \approx 2$.  For the normalized frequency of the bonded mode ($a_{m}/\lambda_{c} \approx 0.4$), this corresponds to $L_{\text{OM}} \approx 0.8 \lambda_{c}$, as expected from the arguments laid out in the introduction.  The TE$_{-,0}$ has a significantly smaller optomechanical coupling due to its reduced electric field energy in the slot.  

An estimate of the optomechanical coupling to the different in-plane differential mechanical modes can be approximated by averaging the displacement amplitude field pattern of the mechanical mode against the (normalized) optical intensity pattern of the zipper cavity optical modes\cite{ref:Johnson5}.  Given the odd symmetry of the even-order $h_{qd}$ mechanical modes, and the even symmetry of the optical intensity for the zipper cavity optical modes, the optomechanical coupling to the even-order $h_{qd}$ modes is approximately zero.  The odd-order in-plane differential modes, on the otherhand, have an anti-node of mechanical displacement at the optical cavity center and an even long-axis symmetry.  For mode numbers small enough that the half-wavelength of the mechanical mode is roughly as large, or larger, than the effective length of the optical cavity mode along the axis of the beam, the optical mode will only sense the central half-wave displacement of the mechanical mode and the optomechanical coupling should still be quite large.  As an example, from the intensity plot of the TE$_{+,0}$ fundamental bonded optical mode in Fig. \ref{fig:optical_mode_profiles}, the effective length of the optical mode along the long-axis of the nanobeams is $L_{\text{eff}} \sim 7 a_{m} = 4.2$ $\mu$m (for $\lambda_{c}=1.5$ $\mu$m).  The mechanical mode index $q$ is roughly equal to the number of half-wavelengths of the mechanical mode along the axis of the zipper cavity.  Therefore, for the zipper cavity of length $l=36$ $\mu$m studied above, the $9$th order in-plane differential mechanical mode with $\Omega_{h_{9d}}/2\pi \approx 170$ MHz has a half-wavelength equal to $4$ $\mu$m $\approx L_{\text{eff}}$.  The resulting optomechanical coupling of the $h_{9d}$ mechanical mode to the TE$_{+,0}$ optical mode is then still relatively large, equal to approximately half that of the coupling to the fundamental $h_{1d}$ mechanical mode.  The optomechanical coupling factor to the TE$_{+,0}$ for each of the in-plane differential modes is tabulated along with the mechanical mode properties in Table \ref{tab:mech_mode_properties}.

There are several physical ways of understanding the strength of the optomechanical coupling represented by $g_{\text{OM}}$.  The per-photon mechanical force is given by $F_{ph} = \hbar g_{\text{OM}}$.  For the zipper cavity, this yields a per-photon force of $F_{ph} \approx \hbar \omega_{c}/\lambda_{c}$, which at near-infrared wavelengths corresponds to $0.2$ pN/photon.  Such a force could be measured using other, non-optical techniques, and could provide a means for detecting single photons in a non-demolition manner.  Also, through the optomechanical coupling, intra-cavity light can stiffen\cite{ref:Meystre,ref:Corbitt1,ref:Hossein_Zadeh1,ref:Sheard1} and dampen\cite{ref:Arcizet,ref:Gigan1,ref:Schliesser2,ref:Corbitt1,ref:ThompsonJD1,ref:Kippenberg4} the motion of the coupled mechanical oscillator\cite{KippenbergOE}.  A perturbative analysis shows that in the sideband unresolved limit ($\Omega_{M} \ll \Gamma$) the effective mechanical frequency ($\Omega_{M}^{\prime}$) and damping rate ($\gamma_{M}^{\prime}$) are given by the following relations (see Ref. \cite{KippenbergOE}):

\begin{align}
(\Omega_{M}^{\prime})^2 = \Omega_{M}^2 + \left(\frac{2|a_{0}|^2g_{\text{OM}}^2}{\Delta^2\omega_{c}m_{u}}\right)\Delta_{o}^{\prime}, \label{eq:ren_Omega_M_2} \\ 
\gamma_{M}^{\prime} = \gamma_{M} - \left(\frac{4|a_{0}|^2g_{\text{OM}}^2\Gamma}{\Delta^4\omega_{c}m_{u}}\right)\Delta_{o}^{\prime}, \label{eq:ren_gamma_M_2}
\end{align}

\noindent where $\Omega_{M}$ and $\gamma_{M}$ are the bare mechanical properties of the zipper cavity, $|a_{0}|^2$ is the time-averaged stored optical cavity energy, $\Delta_{o}^{\prime} \equiv \omega_{l} - \omega_{c}$ is the laser-cavity detuning, $\Gamma$ is the waveguide-loaded optical cavity energy decay rate, and $\Delta^2 \equiv (\Delta_{o}^{\prime})^2 + (\Gamma/2)^2$.  The maxmimum ``optical spring'' effect occurs at a detuning point of $\Delta_{o}^{\prime} = \Gamma/2$.  For this laser-cavity detuning, a single cavity photon introduces a shift in the mechanical frequency corresponding to $\Delta(\Omega_{M}^2)/\Omega_{M}^2 = (2 Q \hbar\omega_{c})/(\lambda_{c}^2 k_{\text{eff}})$.  For the zipper cavity, with $Q = 5 \times 10^6$, $\lambda_{c}=1$ $\mu$m, and $k_{h_{1d}} = 100$ N/m, the resulting single-photon mechanical frequency shift is approximately $\Delta(\Omega_{M})/\Omega_{M} = 1 \%$.  Thus, even a single cavity photon may yield a measurable shift in the mechanical resonance frequency.     

\section{Summary and Discussion}

Using photonic crystal concepts, we have designed an optomechanical system in which optical and mechanical energy can be co-localized in a cubic-micron volume and efficiently coupled through the gradient optical force.  In the particular design studied here, a ``zipper'' cavity consisting of two nanoscale beams of silicon nitride, doubly clamped and patterned with a linear array of air holes, is used to form the optical cavity and the mechanical resonator.  Mechanical oscillations consisting of differential motion of the doubly-clamped silicon nitride nanobeams results in an optomechanical coupling constant as large as $g_{\text{OM}} \sim \omega_{c}/\lambda_{c}$, where $\omega_{c}$ and $\lambda_{c}$ are the optical resonant cavity frequency and wavelength, respectively. This coupling is several orders of magnitude larger than has been demonstrated in high-Finesse Fabry-Perot cavities, and is more than order of magnitude larger than for whispering-gallery micrototoid structures, both of which rely upon the radiation pressure force.  Finite-element-method (FEM) simulations of the zipper cavity show that a structure with an optical $Q = 5 \times 10^6$, mechanical resonance frequency of $\Omega_{M}/2\pi \approx 170$ MHz, and motional mass of $m_{u} \approx 40$ picograms is possible.  In the future, further increase in the mechanical frequency and reduction in the motional mass may be attained by using planar phononic crystals\cite{ref:OlssonIII1} to form the mechanical resonator.  The combination of phononic and photonic crystals would also provide an integrated, chip-scale platform for routing and coupling optical and mechanical energy.

Beyond cavity optomechanics, the zipper cavity may also find application in the field of cavity QED.  In particular, the zipper cavity as described here is suitable for a broad range of wavelengths from the visible to the mid-infrared.  The optical mode volume is made smaller by the sub-wavelength slot gap between the nanobeams\cite{ref:Robinson1}, with $V_{\text{eff}} \sim 0.2 (\lambda_{c}^3)$ for a slot gap of $s \sim \lambda_{c}/10$.  As an example, one can imagine placing nanoparticles of diamond (a popular solid-state system for quantum information processing\cite{ref:Gruber1,ref:Santori2,ref:Jelezko1,ref:Dutt1,ref:Childress1,ref:Beveratos2}) in the gap between the nanobeams.  Such ``pick-and-place'' techniques have been used with other, larger, optical cavities with good success\cite{ref:Kuhn1,ref:Schietinger1,ref:Barclay_NV}.  In the zipper cavity case, the small $V_{\text{eff}}$ would produce a coherent coupling rate with the zero-phonon-line (ZPL) of the NV$^{-}$ transition of approximately, $g_{\text{ZPL}}/2\pi \sim 3$ GHz, even after accounting for the $3-5 \%$ branching ratio for the ZPL line.  This is more than 100 times the radiative-limited linewidth measured for the NV$^{-}$ transition ($12$ MHz), and more than 10 times the theoretical zipper cavity decay rate ($90$ MHz), putting the coupled system deep within the strong coupling regime.  The additional benefit provided by the zipper cavity is the ability to rapidly tune the cavity frequency into and out of resonance with the ZPL of the NV$^{-}$ transition.  If mechanical resonance frequencies could be increased towards GHz values, using the suggested phononic crystal concepts for instance, then new approaches to photon-mediated quantum interactions and quantum state transfer can be envisioned for solid-state cavity QED systems.  

\section*{Acknowledgements}

The authors would like to thank Qiang Lin for extensiive discussions regarding the calculation of motional mass and optomechanical coupling.  The authors would also like to thank Patrick Herring for some of the initial ideas regarding gradient force optomechanical structures.  This work was supported by a DARPA seed grant, managed by Prof. Henryk Temkin, and a NSF EMT grant.
 
\end{document}